%% file: Lietal16t-ase.tex
\documentclass[paper]{IEEEtran}
\usepackage{mathptmx} 
\usepackage{times} 
\usepackage{amsmath} 
\usepackage{amssymb}  
\usepackage{psfrag}
\usepackage{pmat}
\usepackage[table]{xcolor}
\usepackage{cite}
\usepackage{algorithmic}
\usepackage{array}
\usepackage{stfloats}
\usepackage{graphicx}
\usepackage{subfigure} 
\usepackage{algorithm}
\usepackage{epsfig} 
\usepackage{mathptmx} 
\usepackage{times} 
\usepackage{amsmath} 
\usepackage{amssymb}  
\usepackage{float} 
\usepackage{footnote}
\makesavenoteenv{tabular}
\usepackage{threeparttable}

\newtheorem{theorem}{Theorem}

\newtheorem{remark}{Remark}

\usepackage{verbatim,pifont,color,amssymb,amsfonts,amsmath,graphicx,pgf,slashbox,psfrag,ifpdf}
\usepackage[mathscr]{eucal}

\def\defeq{\triangleq}
\def\T{\mbox{\tiny T}}
\def\N{\mbox{\tiny N}}

\def\-1{\mbox{\tiny -1}}

\def\tcr{\textcolor{red}}

\begin{document}
%
\title{Parameter estimation for linear control valve with hysteresis}

\author{{\large Li~Liang~\IEEEmembership{Member,~IEEE,}
        Liu~Jiannan,~\IEEEmembership{Student Member,~IEEE,}\\
         and ~Wan~Huaqing}
\thanks{Li Liang, Liu Jiannan and Wan Huaqing are with the National Lab of Science and Technology on Steam Power for NS, and Wuhan Second Ship Design and Research Institute, Wuhan 430064, China. (e-mail: palmerds@163.com).}
\thanks{This work was supported in part by the Natural Science Foundation of China under Grants 51406138, 61374123.}
}
\markboth{Submitted to IEEE Transactions on Automation Science and Engineering,~May~2016}%
{Liang~Li \MakeLowercase{\textit{et al.}}: Parameter estimation for linear control valve with hysteresis}

\maketitle

\begin{abstract}
The problem of estimating parameters of linear control valve with hysteresis is considered. The hysteretic behavior of control valve is formulated as a switched linear model. An indicator vector, which shows the switching epochs of switched linear model, is explored by subspace decomposition on measurements. With the help of indicator vector, the noisy measurements are classified into separate groups, each corresponding to up-stroke and down-stroke of control valve respectively. The least squares technique is adopted to estimate the parameters of control valve. It is shown that the proposed technique exactly estimates the parameters and switching epochs in absence of noise and exhibits dominant advantage in noisy case.
 \end{abstract}


\vspace{0.5em}

 \ifCLASSOPTIONpeerreview
 \begin{center} \bfseries EDICS Category: 3-BBND \tcr{Check on this.} \end{center}
 \fi
%
\IEEEpeerreviewmaketitle

\section{Introduction\label{Sec:1}}
\begin{figure}
  \centering
  \includegraphics[width=0.7\hsize]{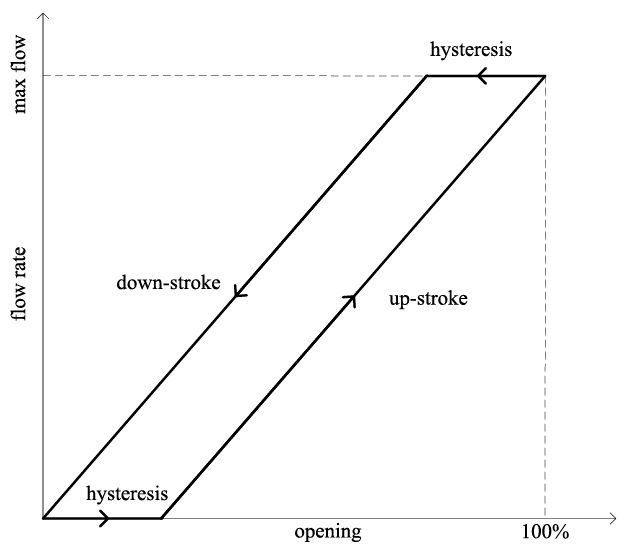}
  \caption{A sketch for the relationship between flow rate and opening for linear control valve with hysteresis. The operating pressure is assumed to be constant.}\label{fig_hysteresis}
\end{figure}

We consider the problem of estimating parameters of linear control valve with hysteresis. Control valve adjusts the flow rate via changing its opening or cross section. Given the operating pressure, the relationship between flow rate and opening totally determines the performance of control valve\cite{Headley03,Darling11}. For an ideal linear control valve, the function from  the opening to flow rate is linear and injective, usually parameterized by the flow coefficient $C_v$. However, the misalignment of installing, screw loose and abrasion of valve disc or injection components all results in a hysteresis to control valve. As illustrated in Fig.\ref{fig_hysteresis}, a hysteretic linear control's behavior switches between two modes which correspond to up-stroke and down-stroke respectively.  Under this case, the parameters of linear control valve, namely, the flow coefficient $C_v$  and the hysteresis, as well as the switching epochs of strokes are unknown quantities to be estimated.

Thus, for a linear control valve involving hysteresis, we consider the following parameter estimating problem: given a set of noisy measurements of opening and flow rate $\{\mu_n, q_n\}, n=1,\ldots, N$, estimate the flow coefficient, the hysteresis and the set of switching epochs. Other quantities of control valve are either assumed known or treated as nuisance parameters.

We note that the estimating problem above is one of the mixed integer type: the flow coefficient and the hysteresis are continuous, and the switching epochs are discrete. The lake of continuity in the parameter space make it difficult to carry out the estimation directly. If we are able to associate each measurement $(\mu_n, q_n)$ with a particular mode, i.e., up-stroke or down-stroke, then we essentially get a classical parameter estimation problem.

As a typical actuator, control valve plays an indispensable role in industry control field\cite{Desborough:01CPC}. The quality of mathematical model of control valve directly relates to the performance of control loop\cite{Choudhury:07Springer}. Fault diagnosis and state monitoring for control valve in some degree also require a precise model\cite{Bartys:06CEP,Hagglund:11CEP}. One of the earliest work of modeling the control valve is by D. McCloy and R. H. Mcguigan, where just the valve coefficient is considered\cite{McCloy64}. There exist two kinds of techniques to model the behavior of control valve, i.e., computation fluid dynamic (CFD) model\cite{Morris:89JFE,Holger06,Glenn08} and standard window model\cite{Fei:11JPC}. CFD model usually applies to describe flow field distribution; the standard window model states the flow property of control valve in an input-output way. The former technique investigates the effects of structure of the control valve, for example the cross section, and is employed in the design period. The standard window model traditionally describes the quantitative relationship between flow rate and cross section. This paper is dedicated to the later technique, which is more applicable in control loop design and fault diagnosis.

Based on the shape of cross section, the standard window models for control valve consists of linear ones, equal percentage ones and fast opening ones. In ideal condition, the characteristics of control valve is a one-one map between flow rate and opening, and its mathematical model has been well studied\cite{Jelali:07JPC}. In such case, the problem of parameter estimation of control valve essentially boils down to one of scaler or point estimation, since there is only one unknown parameter, i.e., the flow coefficient $C_v$.  Unfortunately, this ideal condition usually is destroyed by non-linearity, hysteresis and stiction. Standard window model frequently fails to calculate the flow rate in applications of high-precision control and the incipient fault detection. Thus, it's significantly important to consider these non-ideal factors in standard window model.

When disturbed by a hysteresis, control valve's behavior splits into two modes: the up-stroke and the down-stroke. The basic problem is to attach the measurements with a particular mode. Intuitively, the stroke switchings of control valve will reflect on the time series of openings. However, when the measurements are corrupted by noise or time series information is missed, it's difficult to tell the stroke switchings, and classifications of the measurements to their real modes are no longer definite.

Based on this understanding, the linear control valve with hysteresis is formulated as a switched linear models (SLMs). There has been substantial literatures for parameter estimation of SLMs\cite{Branichy,Heemels}, and some sophisticated techniques are proposed to estimate its parameters, such as clustering-based ones\cite{Ferrari,HayatoNakada}, statistical inference based ones\cite{Bemporad2005,Juloski2} algebraic geometric based ones\cite{Vidal,Roll,Baric_2008,Feng_C}. As a special type of switched linear models, the linear control valve possesses a particular subspace structure in its measurements, which is exploited by singular value decomposition (SVD) technique in this paper. This subspace structure indicates the mode or stroke switchings of control valve. Then clustering technique on subspace structure is employed to associate the measurements to particular mode, as well as to figure out the mode switching epochs. Once the measurements are classified, the parameters of control valve can be estimated by traditional techniques, such as least squares. The proposed technique is a combination of subspace-based technique and clustering-based technique. As a special data structure is utilized, the proposed technique exhibits an improved performance compared with the existing techniques. In the absence of noise, the proposed technique is expected to calculate the exact parameters of linear control valve with feasible computation. When the measurements are corrupted by noise, the proposed technique is capable to estimate the parameters more robustly and numerically stably.

The remainder of the paper is organized as follows: Section \ref{Sec:2} presents a switched linear model for control valve with hysteresis. Here we presents the set of assumptions made in this paper and discuss their implications. Section \ref{Sec:3} presents the main result on the special structure on subspace of the measurements, and an indicator vector is introduced to show the stroke switchings of control valve. Section \ref{Sec:4} presents a technique to estimate the indicator vectors from the noisy measurements, then proposes a dual-SVD based algorithm to classify the measurements and estimate the parameters of control valve in a integrated least squares form. Section \ref{Sec:5} presents a case study on real data and some comparisons with existing techniques. The last section \ref{Sec:6} gives the conclusions.
\section{Problem formulation\label{Sec:2}}
\input ProblemFormulation_v1

\section{Subspace decomposition on meansurements\label{Sec:3}}
\input SubspaceDecompostion_v1

\section{Spectral clustering technique on subspace\label{Sec:4}}
\input SpectralClustering_v1
\section{Case study\label{Sec:5}}
\input Simu_v1

\section{Conclusions\label{Sec:6}}
This paper presents a subspace decomposition technique to the identification of linear control valve with hysteresis. The main contribution is the idea of exploring the particular eigenstructure of $VV^{\T}$, that indicates the modes of control valve.  Compared with the existing techniques, the proposed technique exhibits stable numerical behavior and performance advantages in the case study, thanks to the use of SVD.

A number of practical issues are not discussed in the current paper, most significant is the performance analysis for the proposed technique under different signal-noise-ratio (SNR) levels. It is expected that the performance of the proposed technique tends to degenerate in lower SNR scenarios, as its misclassification ratio rises. Theoretically, there is ultimate performance index,  similar to the Cram\'{e}r-Rao bound. Moreover, the outlier-removing techniques are not considered in this paper as there are some fake measurements, which deteriorates performance of the algorithm, during the stroke-switching process.


\bibliographystyle{IEEEtran}
\bibliography{ref}


\vspace*{-2\baselineskip}
\begin{IEEEbiography}
[{\includegraphics[width=1in,height=1.25in,clip,keepaspectratio]{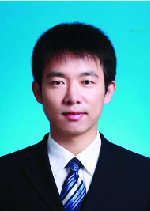}}]{Li Liang}
(S'11,M'13) received the B.E. degree in Department of Automation from Xi'an JiaoTong University, Xi'an, China, in 2007, and B.E. and Ph.D. degrees from the Department of Automation, Tsinghua University, Beijing, China, in 2010 and 2013, respectively. He is currently a researcher in the National Lab of Science and Technology on Steam Power for NS, Wuhan, China. His main research interests include optimal control and estimation theory of switched and hybrid systems.\newline E-mail: palmerds@163.com.
\end{IEEEbiography}

\end{document}

%% file: ProblemFormulation_v1.tex
In this section, we formulate the linear control valve into a switched linear model, as illustrated in Fig.\ref{fig_hysteresis}, and presents main assumptions made in this paper.

A static flow characteristic of control valve can be defined as follows
\begin{equation}\label{eq_Qs}
q=C_vS(\mu)\sqrt{p^2_{in}-p^2_{out}},
\end{equation}
where $S(\mu)$ is the flow area or cross section, which is a function of opening $\mu$; $p_{in},p_{out}$ are the pressures before and after the valve, and differential pressure of valve is defined as $\Delta p=p_{in}-p_{out}$; $C_v$  refers to flow coefficient, which indicates the flow capability of the valve.

In ideal condition, the flow area function $S(\mu)$ is injective. Based on the function form, the control valve is typically divided into three types: linear ones with linear function, equal-percentage ones with exponential function and fast-open ones with power function. When the control valve is distorted by hysteresis, the flow area function changes into a multi-value one, which means the same opening $\mu$ in up-stroke and down-stroke causes different flow rates $q$s. Define a switched linear function for the flow area as follows
\begin{equation}
    S(\mu)= \left\{\begin{array}{ll}
    \alpha \mu , &  \mu \text{ in down-stroke}\\
    \alpha \mu + \beta, &  \mu \text{ in up-stroke}\\
    \end{array} \right.,  \label{eq:windows}
\end{equation}
where the up-stroke and down-stroke have the same slope coefficient, i.e., $\alpha$, and up-stroke has a intercept $\beta$ which models the hysteresis of valve.


For the concision of expression, we defined following measurement matrices:
\begin{eqnarray}
x_n &\triangleq & \mu_n,~~~X   \triangleq \left[x_1, \ldots, x_N\right],\label{eq:x}\\
y_n &\triangleq & \frac{q_n}{C_v\sqrt{p^2_{in,n}-p^2_{out,n}}},~~~Y   \triangleq \left[y_1, \ldots, y_N\right].\label{eq:y}
\end{eqnarray}
Then, the flow characteristic of linear control can be formulated by a switched linear models
\begin{eqnarray}
    y_n &=& \left\{\begin{array}{ll}
    \alpha x_n, &  n \in \ell_{down}\\
    \alpha x_n+\beta, &  n \in \ell_{up}\\
    \end{array} \right., \label{eq:SLS}
\end{eqnarray}
where $\ell_{down}, \ell_{up}$ denote the label sets of down-stroke and up-stroke, in which the flow area functions are parameterized by coefficient $\alpha$ and $\alpha, \beta$ respectively.

Thus, the estimating problem can be formally stated as: given the input-output measurements matrices $\{X, Y\}$, estimate the parameters $\alpha, \beta$ and the mode sets $\ell_{up}, \ell_{down}$.


Main assumptions made in developing our technique are as follows:
\begin{itemize}
\item[A1:] the measurements are corrupted by gaussian noise and its statistic characteristics are unknown.

\item[A2:] the measurements represents the static and steady characteristics of valve. The dynamic characteristic is not considered in this paper. This assumption is reasonable, as flow distribution of valve usually reaches a steady state within a very short period. So the dynamic characteristic of valve can be ignored when investigating its input-output properties.

\item[A3:] the flow is considered as incompressible fluid. The properties of compressible flow  are beyond the description capability of standard window model, which is usually analysed by CFD model.

\item[A4:] We consider the off-line estimating problem of control valve, and the sampling period is assumed wider than the enduring time of mode switching, which means the time-series information is unavailable.

\end{itemize}

%% file: SubspaceDecompostion_v1.tex
In this section, we explore the special structure in the subspace of measurements, and construct an indicator vector which exactly shows the mode switchings of control valve in the absence of noise.

To develop the main result, we reformulate the switched linear models in (\ref{eq:SLS}) in a unified form as follows:
\begin{equation} \label{eq:Z=AU}
Z=ADP,
\end{equation}
where $P$ is an unknown permutation matrix that permutes the ordered opening sequence to the actual sequence of arrivals, parameter matrix $A$ and opening matrix $D$ are defined as
\begin{eqnarray}
A &\defeq & \left[\begin{array}{cc}
  1 & 0\\
    \alpha &\beta\\
  \end{array}\right], \\
D & \defeq &  \left[\begin{array}{cccccc}
    d_1^{(1)}&\ldots&d_{N_1}^{(1)}&d_{1}^{(2)}&\ldots& d_{N_2}^{(2)}\\
    0&\ldots&0&1&\ldots& 1
  \end{array}\right]\nonumber\\
&=&  \left[\begin{array}{cc}
    D_1&D_2\\
    \mathbb{O}^{\T}&\mathbb{I}^{\T}
  \end{array}\right].  \label{eq:diagU}
\end{eqnarray}
$A$ is the parameter matrix which totally describes the characteristics of linear control valve; $D$ is the ordered opening matrix where $D_1=[d_1^{(1)},\cdots, d_{N_1}^{(1)}]$ denotes the $N_1$  openings from down-stroke mode and $D_2=[d_1^{(2)},\cdots, d_{N_2}^{(2)}]$ denotes the $N_2=N-N_1$ openings from up-stroke mode. $\mathbb{O}_{N_1\times 1}$ and $\mathbb{I}_{N_2\times 1}$ are all-zero and all-one column vectors with compatible dimension.

The following theorem captures the special structure in the row space of data matrix $Z$.
\begin{theorem}\label{thm1}
  Let a noiseless measurement matrix $Z$ in $(\ref{eq:Z=AU})$ have the singular value decomposition of the form
  \begin{equation}
    \label{SVD}
    Z=Q\Sigma V^{\T},
  \end{equation}
  where $QQ^T=I$ and $V^{\T}$ of the same size as $D$ with orthogonal rows.

  Define an indicator vector
   \begin{equation}
   \label{indicate}
    H\defeq P^{\T}\left[\begin{array}{c}
    \mathbb{O}\\
    \mathbb{I}
  \end{array}\right],
  \end{equation}
  where $P$ is an unknown permutation matrix, and $\mathbb{O}_{N_1\times 1}$ and $\mathbb{I}_{N_2\times 1}$ are all-zero and all-one column vectors with compatible dimension.

  Then the indicating vector $H$ is a right eigenvector of $VV^{\T}$, i.e.,
  \begin{equation}
  VV^{\T}H=H.
  \end{equation}
\end{theorem}

\vspace{0.5em}
\begin{IEEEproof}
Let $\tilde{Z}=AD$ denote the ideal case where samples of down-stroke and up-stroke modes arrive sequentially, i.e., $P=I$, then $\tilde{Z}$ has the singular value decomposition of the form
  \begin{equation}
  \tilde{Z}=\tilde{Q}\tilde{\Sigma} \tilde{V}^{\T},
  \end{equation}
Note that $\tilde{\Sigma}$ is a diagonal matrix with reduced dimension $2\times 2$, $\tilde{Q}$ is an orthogonal matrix, and $\tilde{V}^{\T}$ is a $2\times N$ matrix of the same dimension of input matrix $D$. As parameter matrix $A$ is nonsingular, $\tilde{V}^{\T}$ spans the same row space as $D$ does. In other words, we have
  \begin{equation}
    \tilde{V}^{\T}=TD, T\defeq \tilde{\Sigma}^{\-1}\tilde{Q}^{\T}A.
  \end{equation}
Because $VV^{\T}=TDD^{\T}T^{\T}=I$ and $T$ is nonsingular, thus
  \begin{equation}
    T^{\T}T=\left(DD^{\T}\right)^{\-1}.
  \end{equation}
Then we have
\begin{eqnarray}
  \tilde{V}\tilde{V}^{\T} &=& D^{\T}T^{\T}TD \nonumber\\
   &=& D^{\T}\left(DD^{\T}\right)^{\-1}D \nonumber\\
   &=& D^{\T}\left[
                \begin{array}{cc}
                  e_1+e_2 & e_2 \\
                  e_2 & N_2 \\
                \end{array}
              \right]^{\-1}D   \nonumber\\
   &=& \frac{1}{\left|DD^{\T}\right|}D^{\T}\left[
                \begin{array}{cc}
                  N_2 & -e_2 \\
                  -e_2 & e_1+e_2 \\
                \end{array}
              \right]D
\end{eqnarray}
where $e_1\defeq \sum_{i=1}^{N_1}d_i^{(1)},e_2\defeq \sum_{i=1}^{N_2}d_i^{(2)}$.

Therefor, we have
  \begin{eqnarray}
    \tilde{V}\tilde{V}^{\T}\left[\begin{array}{c}
                \mathbb{O}\\
                \mathbb{I}
                \end{array}\right]
   &=& \frac{1}{\left|DD^{\T}\right|}D^{\T}\left[
                \begin{array}{cc}
                  N_2 & -e_2 \\
                  -e_2 & e_1+e_2 \\
                \end{array}
              \right]\left[\begin{array}{c}
                                e_2\\
                                N_2
                \end{array}\right]  \nonumber\\
   &=& \frac{1}{\left|DD^{\T}\right|}D^{\T}\left[\begin{array}{c}
                                0\\
                                -e_2^2+N_2(e_1+e_2)
                \end{array}\right] \nonumber\\
   &=& \frac{-e_2^2+N_2(e_1+e_2)}{\left|DD^{\T}\right|}\left[\begin{array}{c}
                \mathbb{O}\\
                \mathbb{I}
                \end{array}\right] \nonumber\\
   &=& \left[\begin{array}{c}
                \mathbb{O}\\
                \mathbb{I}
                \end{array}\right] \label{eq:VVT}
\end{eqnarray}
For the permutated measurement matrix $Z=\tilde{Z}P$, we have $Q=\tilde{Q}$, $\Sigma=\tilde{\Sigma}$ and $V^{\T}=\tilde{V}^{\T}P$. As $P$ is a permutation matrix, $PP=I$, so we have
  \begin{equation}
    \tilde{V}^{\T}=V^{\T}P \label{eq:tilde_V}.
  \end{equation}
Substitute (\ref{eq:tilde_V}) into (\ref{eq:VVT}), and the proof is completed.
\end{IEEEproof}

\vspace{0.5em}


If the opening sequence arriving in consecutive from the same mode, we have $P=I$.  In this case, matrix $VV^{\T}$ has a eigenvector in form of $[\mathbb{O}^{\T},\mathbb{I}^{\T}]^{\T}$, where zeros indicate the positions of down-stroke mode and ones indicate the ones of up-stroke mode. While matrix $P$ in general scrambles the structure of $[\mathbb{O}^{\T},\mathbb{I}^{\T}]^{\T}$, the permutated vector $P^{\T}[\mathbb{O}^{\T},\mathbb{I}^{\T}]^{\T}$, referred as indicator vector $H$, can be treated as indicator for the stroke modes of valve. The problem of classifying the measurements is turned into one of classifying the elements of indicator vector. For noiseless measurement, the elements of indicator vector can be easily classified via a threshold; When the measurements are distorted by noise, the elements of indicator vector can be classified by clustering on scalars.

%% file: SpectralClustering_v1.tex
In this section, we dedicate to draw the indicator vector from the noisy measurements, and then propose an algorithm to solve the parameter estimating problem of control linear valve.

Theorem~\ref{thm1} reveals the indicator vector is one of eigenvectors of $VV^{\T}$. We are not sure if the indicator vector locates among the standard eigenvectors of $VV^{\T}$ calculated by SVD directly; if it does, its exact position is unfixed as there are two eigenvectors for $VV^{\T}$; if it doesn't, it must be a linear combination of standard eigenvectors of $VV^{\T}$. Fortunately, there are only two eigenvectors for $VV^{\T}$, which correspond to the first two columns of right eigenmatrix.

Let $VV^{\T}$ has a SVD of the form
\begin{equation}\label{eq:W}
VV^{\T}=WGW^{\T},
\end{equation}
where $W=[w_1,w_2]$ is a $N\times 2$ matrix with its columns spanning the eigenvector space of $VV^{\T}$, namely $H=u_1w_1+u_2w_2$. Thus, the indicating vector can be expressed as a linear combination of $W$, i.e.,
\begin{equation}\label{eq:W}
H=WU,U\defeq\left[u_1,u_2\right]^{\T}.
\end{equation}

The remained problem is turned into estimating the weight vector $U$. Suppose there is a small amount of, at least two, pre-classified measurements in $Z$, denote their label set as $\tilde{\ell}$ and data set as $Z_{\tilde{\ell}}$, we can draw a matrix $\hat{W}_{\tilde{\ell}}$ from $W$ at the corresponding columns, and create an indicating vector $\tilde{H}_{\tilde{\ell}}$ by $1$ labeling the up-stroke mode and $0$ labeling the down-stroke mode. The weight matrix can be initially estimated as
\begin{equation}\label{eq:iter}
\hat{U}_{\tilde{\ell}}=\left(\tilde{W}^{\T}_{\tilde{\ell}}\tilde{W}_{\tilde{\ell}}\right)^{\-1}\tilde{W}^{\T}_{\tilde{\ell}}\tilde{H}_{\tilde{\ell}}.
\end{equation}
Thus, the indicator vector can be calculated as
\begin{equation}\label{eq:update}
\hat{H}=W\hat{U}_{\tilde{\ell}}.
\end{equation}

Once we get the estimation of indicating vector $\hat{H}$, the measurements $Z$ can be classified by clustering on the elements of $\hat{H}$. Note that, to improve the performance of classification, we can iteratively operate $(\ref{eq:iter})$ and $(\ref{eq:update})$, until getting a stable estimation $\hat{H}$.

Given the classified measurements, the parameters of SLMs can be estimated by standard estimating techniques. Traditionally, the parameters of submodel of SLMs are estimated separately and independently. However, for the case of linear control valve, the separate estimations inevitably result in different parameters, which is inconsistent with the fact that the up-stroke and down-stroke have the same slope $\alpha$. Based on this understanding, we reformulate the linear control model in (\ref{eq:Z=AU}) into an integrated least squares form as follow
\begin{equation}\label{eq:LS}
YP=\left[\alpha, \beta\right]XP,
\end{equation}
where $X, Y$ are defined in (\ref{eq:x}) and (\ref{eq:y}), and $XP$ and $YP$ represent the classified or de-permuted opening and flow rate measurements. The integrated least square form in (\ref{eq:LS}), which avoids inconsistent estimations for slope $\alpha$, can be readily solved by least squares (LS) or total least squares (TLS).

By now, the major ideas of the proposed technique can be divided into three stags: (i) compute the SVDs of the measurement matrix $Z$ and $VV^{\T}$; (ii) estimate the indicator vector $H$ and then apply a clustering technique on it to figure out the label set $\ell_{up}, \ell_{down}$. (iii) estimate the parameters in an integrated least squares form.  The specifics of the implementation is shown in Algorithm~\ref{alg1}.
\begin{algorithm}
  \caption{\label{alg1} The Clustering on the eigenvectors of $VV^{\T}$}
  \begin{algorithmic}
  \REQUIRE \STATE Measurements $X=\{\mu_n\}_{n=1}^{\N},Y=\{y_n\}_{n=1}^{\N}$, pre-labeled set $\tilde{\ell}$
  \ENSURE \STATE Parameter estimations $\{\hat{\alpha},\hat{\beta}\}$, label set $\{\hat{\ell}_{up},\hat{\ell}_{down}\}$
    \STATE (1) compute the SVDs of measurement matrix $Z=[X;Y]$ and $VV^{\T}$:
    \STATE    \qquad\qquad$Z=Q\Sigma V^{\T}$, $VV^{\T}=WGW^{\T},W\defeq[w_1,w_2]$
    \STATE (2) estimate the weight vector $\hat{U}_{\tilde{\ell}}$ and indicating vector $\hat{H}$:
    \STATE     \qquad\qquad$\hat{U}_{\tilde{\ell}}=\left(\tilde{W}^{\T}_{\tilde{\ell}}\tilde{W}_{\tilde{\ell}}\right)^{\tiny -1}\tilde{W}^{\T}_{\tilde{\ell}}\tilde{H}_{\tilde{\ell}}$
    \STATE     \qquad\qquad$\hat{H}=W\hat{U}_{\tilde{\ell}}$
    \STATE (3) cluster the elements of $\hat{H}$ into $2$ groups with $K$-means:
               \STATE\qquad\qquad $[cent, ind]=kmeans(\hat{H},2)$
               \STATE\qquad\qquad $\hat{\ell}_i=find(ind==i), i=1,2$,
               \STATE where the label set $\hat{\ell}_1,\hat{\ell}_2$ can readily map to $\hat{\ell}_{up},\hat{\ell}_{down}$  by the cluster centroid matrix $cent$.
    \STATE (4) estimate the parameters via least squares:
            \vspace{-0.5em}
            \begin{eqnarray}
                [\hat{\alpha},\hat{\beta}] &= & \tilde{Y}\tilde{X}^{\T}(\tilde{X}\tilde{X}^{\T})^{\mbox{\tiny -1}}, \nonumber\\
                \tilde{X} & \defeq &  \left[ \begin{array}{cc}
                                          X(\hat{\ell}_{down}) & X(\hat{\ell}_{up}) \\
                                          \mathbb{O}^{\T}&\mathbb{I}^{\T} \\
                                        \end{array}
                                      \right],\nonumber\\
                \tilde{Y} &\defeq &  [Y(\hat{\ell}_{down}) , Y(\hat{\ell}_{up})]\nonumber
            \end{eqnarray}
            \vspace{-2em}
  \RETURN $\{\hat{\alpha},\hat{\beta}\}$, $\{\hat{\ell}_{up},\hat{\ell}_{down}\}$
\end{algorithmic}
\end{algorithm}

\begin{remark}
In the absence of noise, the eigenvectors $VV^{\T}$ can be used to calculate the indicator vector directly and then associate each measurement to particular mode.  With noise, these eigenvectors no longer contain entries exactly equal to one or zero, we need to adapt the clustering technique to estimate the elements of indicator vector.
\end{remark}

\begin{figure}[thpb]
  \centering
  \includegraphics[width=0.48\hsize]{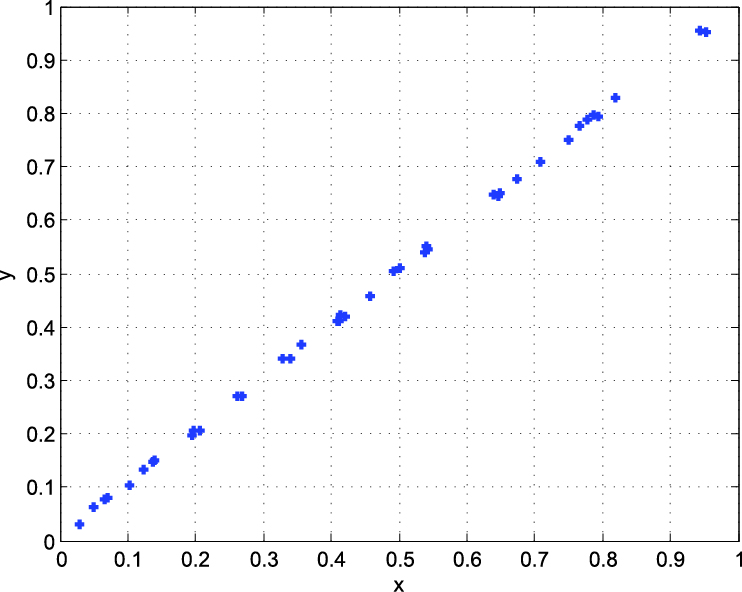}
  \includegraphics[width=0.48\hsize]{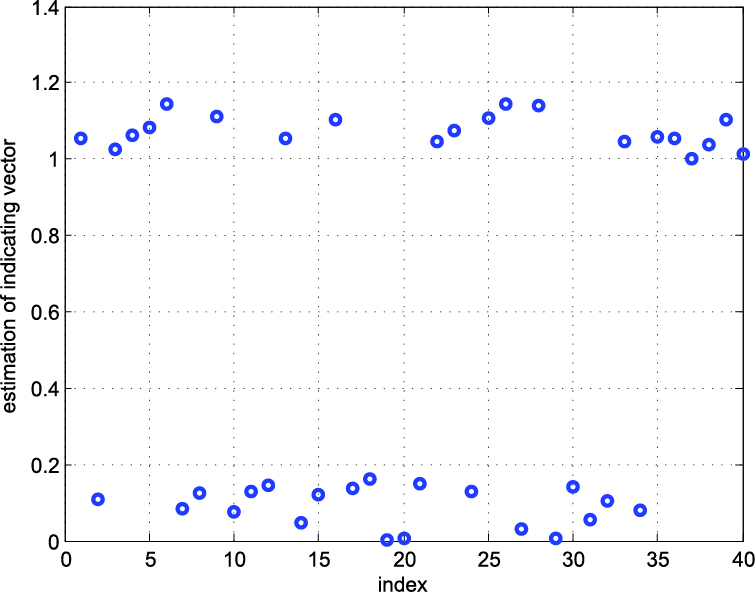}
  \caption{\label{fig:ind} An illustration of the the proposed algorithm. The nominal parameters are $\alpha=1, \beta=0.01$, the signal noise ratio, $SNR=10\log\frac{YY^{\T}}{N\sigma^2}=50$dB, where $\sigma^2$ is the noise variance, the sample number are $N_1=20, N_2=20$ and the number of pre-classified measurements is $2$.}
\end{figure}

Benefited from its algebraic operations, the proposed algorithm is expected to give exact classification of the measurements in the absence of noise, in despite of the magnitude of the hysteresis $\beta$. When the measurements are polluted by noise, the performance of proposed algorithm depends on the estimation of the indicator vector $\hat{H}$, which can be enhanced by increasing the amount of the pre-classified measurements or adopting an iterative technique. Refer to Figure.~\ref{fig:ind} for an illustration of the performance of the proposed algorithm, we can found that the estimation of the indicator vector $\hat{H}$ exhibits a well Clustering characteristics under a small hysteresis.

%% file: Simu_v1.tex
In this section, the proposed technique, referred as svd-based technique, is compared with two benchmark techniques in a numerical study on the real operation data.
The first benchmark technique is directly estimating the parameters of linear control valve, and considering no hysteresis. Intuitively, this is a primary ,and even some kind of rude technique which is taken as a reference here.
The second benchmark, referred as hdc-based technique, is an algebraic geometric technique based on the hybrid decomposition constraints (HDC) in \cite{Vidal,Baric_2008}.  The hdc-based technique embeds the parameters of all modes into a polynomial which is estimated by the least squares, then recovers them through a polynomial differentiation algorithm.

As the nominal parameters remain unknown, we adopt a Relative Fitting Error (RFE) as the performance metric in comparing different techniques, which is defined as follows:
\begin{equation}
    RFE(\hat{y},\hat{y}_0)=\frac{\|\hat{y}-y\|}{\|\hat{y}_0-y\|}
\end{equation}
where $y$ is the actual measurements of flow rate, $\hat{y}$ for the flow predictions of compared technique, $\hat{y}_0$ is the flow predictions of reference technique, which refers to the first technique considering no hysteresis in this paper.

\begin{figure}[thpb]
  \centering
  \includegraphics[width=0.96\hsize]{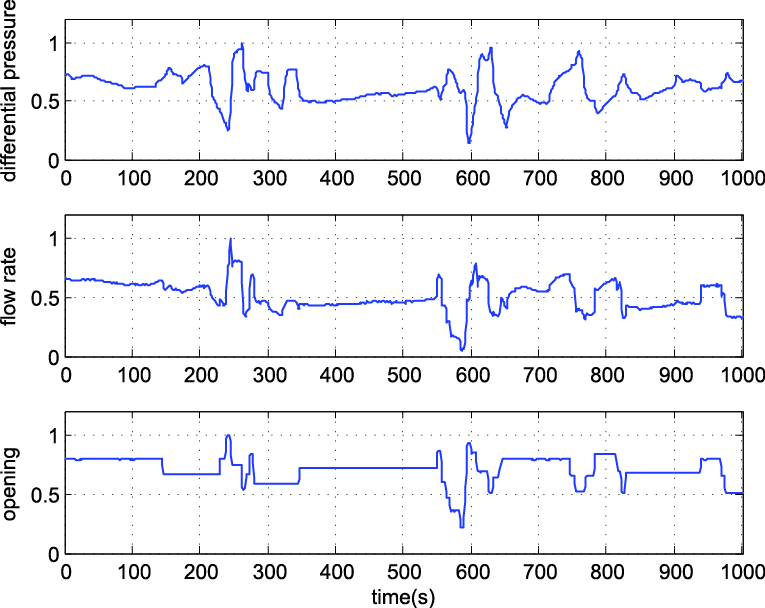}
  \caption{\label{fig:data} Measurements of a linear control valve for steam turbine.}
\end{figure}

Fig.~\ref{fig:data} shows the actual measurements for the estimating parameters, which include the differential pressure $\Delta p_n$, flow rate $q_n$ and the openings of control valve $\mu_n$. The data were distorted by the
measurement noise introduced by the equipments and environment condition. For the sake of protection for industry secret, the data are normalized here, which do not effect the conclusions of case study.

\begin{figure}[thpb]
  \centering
  \subfigure[]{\label{fig:open:a}\includegraphics[width=0.96\hsize]{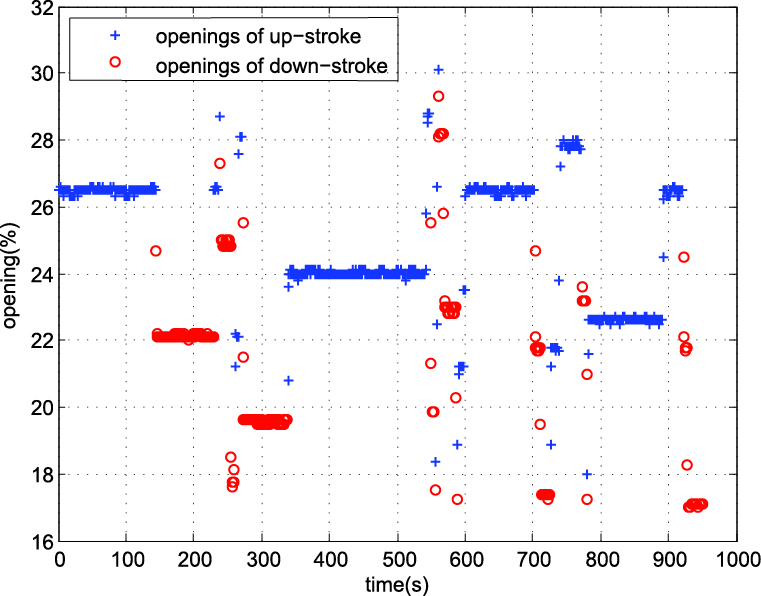}}
  \subfigure[]{\label{fig:open:b}\includegraphics[width=0.96\hsize]{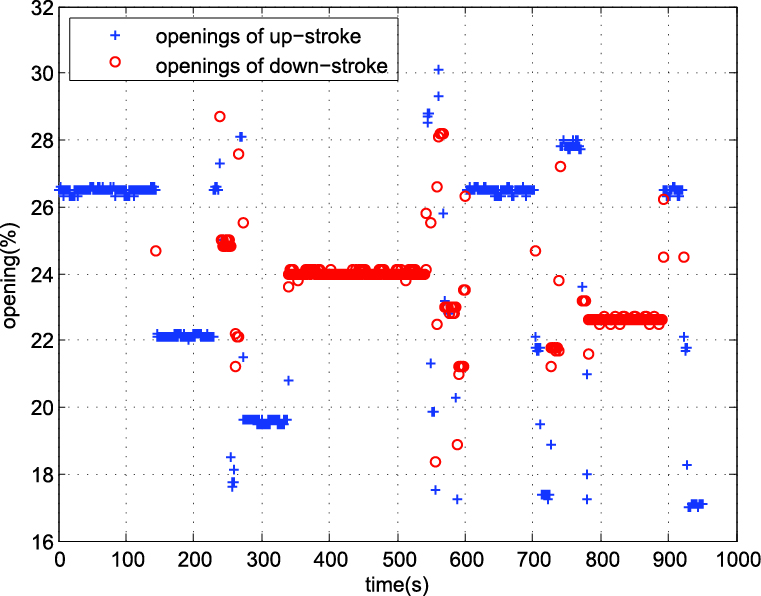}}
  \caption{\label{fig:open}The openings classified for up-stroke and down-stroke modes. (a) for the opening classification of the svd-based technique; (b) for the opening classification of the hdc-based technique.}
\end{figure}

The measurements are classified into up-stroke mode's and down-stroke mode's, refer to Fig.~\ref{fig:open} for the comparison of classification result. The hdc-based technique, as involved nonlinear operations, amplifies the measurement noise and tends to misclassify some short segments of data, refer to Fig.~\ref{fig:open:b} for its openings' classification. In Fig.~\ref{fig:open:a}, the svd-based technique attaches almost all openings to the proper modes, except some outliers which correspond to measurements during the transitional period.

\begin{figure}[thpb]
  \includegraphics[width=0.96\hsize]{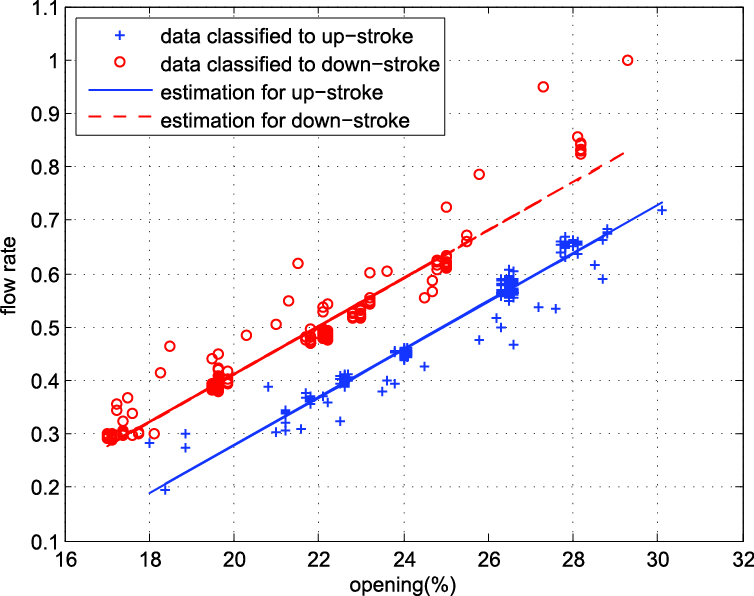}
  \caption{\label{fig:classify}Data classification and estimated model for up-stroke and down-stroke mode. }
\end{figure}

The terminal result of classification and estimation of the svd-based technique is shown in Fig.~\ref{fig:classify}. We should note that it's some kind of improper to classify the outliers of measurements. There outliers essentially provide fake information about the flow property of control valve and contrarily puzzle the classification of measurements. Because, in the transitional period, the flow rate of control valve has no relationship with its opening. Thus, we can forecast that the performance of techniques considered in this paper, including the proposed one, is weakened by there outliers. This effect would be more obvious when more frequent stroke changes are involved.

\begin{figure}[thpb]
  \includegraphics[width=0.96\hsize]{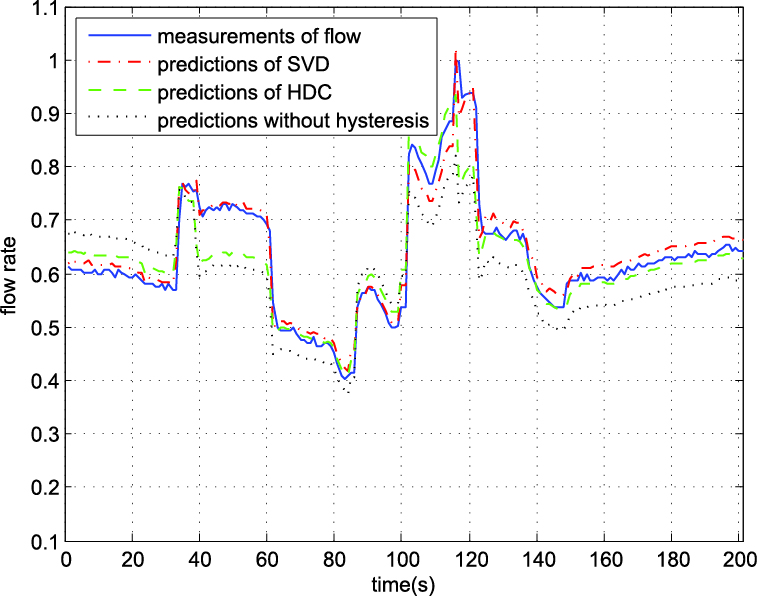}
  \caption{\label{fig:pred}The flow predictions of the proposed technique and benchmark techniques. Taken reference technique's RFE as $RFE_0=1$, the proposed technique, denoted as \textit{SVD}, has a $RFE_{svd}=0.1$, and algebraic geometric technique, denoted as \textit{HDC}, has a $RFE_{hdc}=0.4$.}
\end{figure}

To further verify the performance of the proposed technique, the flow predicting ability of the svd-based technique is compared with benchmarks, over a new batch of measurements. In Fig.~\ref{fig:pred}, We observe that the svd-based technique predicts the flow rate well overall measurements, while the hdc-based technique misses some short segments where fast mode changes occur, for example, the segments between $40$s and $60$s and around $120$s. The reference technique, as considering no hysteresis, mis-predicts almost all flow rates when mode change happens. On the other hand, this also illustrates the destruction that the hysteresis brings to control valve.
